\documentclass[12pt,notitlepage]{amsart}%
\usepackage{amssymb}
\usepackage{amsfonts}
\usepackage{graphicx}
\usepackage{amscd}
\usepackage{graphicx}
\usepackage{amsmath}%
\theoremstyle{plain}

\newtheorem{definition}{Definition}

\newtheorem{notation}{Notation}

\newtheorem{remark}{Remark}

\numberwithin{equation}{section}
\numberwithin{theorem}{section}
\numberwithin{lemma}{section}
\numberwithin{proposition}{section}
\numberwithin{corollary}{section}

\ifx\pdfoutput\relax\let\pdfoutput=\undefined\fi
\newcount\msipdfoutput
\ifx\pdfoutput\undefined\else
\ifcase\pdfoutput\else
\ifx\paperwidth\undefined\else
\ifdim\paperheight=0pt\relax\else\pdfpageheight\paperheight\fi
\ifdim\paperwidth=0pt\relax\else\pdfpagewidth\paperwidth\fi
\fi\fi\fi
\begin{document}
\title[Non-Archimedean Quantum Mechanics]{Non-Archimedean Quantum Mechanics via Quantum Groups}
\author{W. A. Z\'{u}\~{n}iga-Galindo}
\address{University of Texas Rio Grande Valley\\
School of Mathematical \& Statistical Sciences\\
One West University Blvd\\
Brownsville, TX 78520, United States}
\email{wilson.zunigagalindo@utrgv.edu}
\thanks{The author was partially supported by the Debnath Endowed \ Professorship}

\begin{abstract}
We present a new non-Archimedean realization of the Fock representation of the
$q$-oscillator algebras where the creation and annihilation operators act on
complex-valued functions, which are defined on a non-Archimedean local field
of arbitrary characteristic, for instance, the field of $p$-adic numbers. This
new realization implies that many quantum models constructed using
$q$-oscillator algebras are non-Archimedean models, in particular, $p$-adic
quantum models. In this framework, we select a $q$-deformation of the
Heisenberg uncertainty relation and construct the corresponding $q$-deformed
Schr\"{o}dinger equations. In this way we construct a $p$-adic quantum
mechanics which is a $p$-deformed quantum mechanics. We also solve the
time-independent Schr\"{o}dinger equations for the free particle, and a
particle in a non-Archimedean box. In the last case, we show the existence of
a discrete sequence of energy levels. We determine the eigenvalues of
Schr\"{o}dinger operator for a general radial potential. By choosing the
potential in a suitable form we recover the energy levels of the $q$-hydrogen atom.

\end{abstract}
\maketitle

\section{Introduction}

In the last thirty-five years the connection between $q$-oscillator algebras
and quantum physics have been studied intensively, see, e.g.,
\cite{arefeve-volovich}, \cite{Baxter}, \cite{Biedenharn}, \cite{Chung et al},
\cite{Ezran et al}, \cite{Finkelstein-2000}-\cite{Finkelstein-1999},
\cite{Klimyk}, \cite{Lavagno-2010}-\cite{Lavagno}, \cite{Macfarlane},
\cite{Manin}, \cite{Van der Jeugt}, \cite{Wess-Zumino}-\cite{Zhang-2}, and the
references therein. From the seminal work of Biedenharn \cite{Biedenharn} and
Macfarlane \cite{Macfarlane}, it was clear that the $q$-analysis,
\cite{Ernst}, \cite{Kac}, \cite{Klimyk}, plays a central role in the
representation of $q$-oscillator algebras which in turn has a deep physical
meaning. In particular, the $q$-deformation of the Heisenberg algebra drives
naturally to several types of $q$-deformed Schr\"{o}dinger equations, see,
e.g., \cite{arefeve-volovich}, \cite{Lavagno}, \cite{Lavagno2},
\cite{Wess-Zumino}-\cite{Zhang-2}.

Also, in the last thirty years the $p$-adic quantum mechanics and the $p$-adic
Schr\"{o}\-dinger equations have been studied extensively, see, e.g.,
\cite{Albeverio-QM}, \cite{Cianci-QM}-\cite{Dragovivh et al},
\cite{Khrennikov-QM}, \cite{Kochubei-QM}-\cite{Koch}, \cite{Meurice-QM},
\cite{Ruelle-QM}, \cite{Vladimirov-Vol-QM}, \cite{Vourdas-QM}-\cite{V-V-Z},
\cite{Zelenov-QM}, among many available references. Here, we present a new
perspective: a\ non-Archimedean quantum mechanics, which includes $p$-adic
quantum mechanics as a particular case, is a $q$-deformation of the classical
quantum mechanics. The construction\ is based on a new non-Archimedean
realization of the Fock representation of the $q$-oscillator algebras, where
the creation and annihilation operators act on complex-valued functions
defined on a non-Archimedean local field $\mathbb{K}$ of arbitrary characteristic.

On the other hand, the emergence of ultrametricity in physics, which is the
occurrence of ultrametric spaces in physical models, has driven to the
development of deep connections between $p$-adic analysis and physics, see,
e.g., \cite{Dragovivh et al,Vol,Volovich1,Ramal et
al,V-V-Z,Khrennikov,KKZuniga,Varadarajan,Zuniga-LNM-2016} and the references
therein. The existence of a Planck length implies that the spacetime
considered as a topological space is completely disconnected, the points
(which are the connected components) play the role of spacetime quanta. This
is precisely the Volovich conjecture on the non-Archimedean nature of the
spacetime below the Planck scale, \cite{Vol,Volovich1}, \cite[Chapter
6]{Varadarajan}.\ In the last forty years, the above mentioned ideas have
motivated many developments in quantum field theory and string theory, see,
e.g., \cite{Brekke:1993gf}, \cite{Dragovivh et al}, \cite{Freund:1987ck},
\cite{Freund:1987kt}, \cite{Hlousek:1988vu}, \cite{V-V-Z}, \cite{Vol}%
-\cite{Volovich1}, and more recently, \cite{Zuniga-arroyo},
\cite{Zuniga-B-Veys}, \cite{Zuniga-B-Garcia}, \cite{Zuniga-B-Veys},
\cite{Gubser:2016guj}-\cite{Gubser:2017qed}, among others.

In \cite{arefeve-volovich} Aref'eva and Volovich pointed out the existence of
deep analogies between $p$-adic and $q$-analysis, and between $q$-deformed
quantum mechanics and $p$-adic quantum mechanics. In this work we start the
investigation of these matters. We present a new non-Archimedean realization
of the Fock representation of $q$-oscillator algebras, where the creation and
annihilation operators act on functions $f:\mathbb{K}\rightarrow\mathbb{C}$,
where $\mathbb{K}$ is a non-Archimedean local field, for instance, the field
of formal Laurent series:
\[
\mathbb{F}_{q}\left(  \left(  T\right)  \right)  =\left\{
{\displaystyle\sum\limits_{k=k_{0}}^{\infty}}
a_{k}T^{k};a_{k}\in\mathbb{F}_{q}\text{, }a_{k_{0}}\neq0\text{ with }k_{0}%
\in\mathbb{Z}\right\}  ,
\]
where $\mathbb{F}_{q}$ is the finite field with $q$ elements. Our results
imply, for instance, that the results on the $q$-deformed harmonic oscillator
of Biedenharn \cite{Biedenharn} and Macfarlane \cite{Macfarlane} are valid in
$\mathbb{F}_{q}\left(  \left(  T\right)  \right)  $.

We also study some analogues of the Schr\"{o}dinger equation coming from a
$q$-deformation of the Heisenberg algebra. There are several different ways of
choosing a $q$-deformation of the classical uncertainty relation which in turn
produces several different $q$-deformations of the Schr\"{o}dinger equation.
In the case of the free particle, these $q$-deformed equations admit plane
waves which are constructed using the classical exponential functions from
$q$-analysis. There are two basic exponential functions, one admitting a
meromorphic continuation to the complex plane, and the other admitting an
entire analytic continuation to the complex plane. We pick a $q$-deformation
of the uncertainty relation so that the corresponding $q$-deformed
Schr\"{o}dinger equation admits plane waves which are entire functions.

In this framework, we study, in a rigorous mathematical way, the
time-independent $q$-deformed Schr\"{o}dinger equations for a free particle, a
particle confined in a non-Archimedean box, and a particle subject to an
arbitrary radial potential.

We now discuss some applications of our results and do some comparisons. To
fix ideas we discuss all the results using the field $\mathbb{F}_{q}\left(
\left(  T\right)  \right)  $. The standard norm on $\mathbb{F}_{q}\left(
\left(  T\right)  \right)  $ is defined as%
\[
\left\vert x\right\vert =\left\{
\begin{array}
[c]{lll}%
0 & \text{if} & x=0\\
q^{-k_{0}} & \text{if} & x=%
{\displaystyle\sum\limits_{k=k_{0}}^{\infty}}
a_{k}T^{k}\text{, }a_{k_{0}}\neq0.
\end{array}
\right.
\]
Notice that $\left\vert T\right\vert =q^{-1}$. We denote by $S=\left\{
x\in\mathbb{F}_{q}\left(  \left(  T\right)  \right)  ;\left\vert x\right\vert
=1\right\}  $ the unit sphere. Then
\[
\mathbb{F}_{q}\left(  \left(  T\right)  \right)  \smallsetminus\left\{
0\right\}  =%
{\displaystyle\bigsqcup\limits_{k=-\infty}^{\infty}}
T^{k}S.
\]
Which implies that $\mathbb{F}_{q}\left(  \left(  T\right)  \right)
\smallsetminus\left\{  0\right\}  $ is a self-similar set, and that $\left(
\mathbb{Z},+\right)  $ is a scale group acting on $\mathbb{F}_{q}\left(
\left(  T\right)  \right)  \smallsetminus\left\{  0\right\}  $ as
\[%
\begin{array}
[c]{lll}%
\mathbb{Z\times}\left(  \mathbb{F}_{q}\left(  \left(  T\right)  \right)
\smallsetminus\left\{  0\right\}  \right)  & \rightarrow & \mathbb{F}%
_{q}\left(  \left(  T\right)  \right)  \smallsetminus\left\{  0\right\} \\
\left(  k,x\right)  & \rightarrow & T^{k}x.
\end{array}
\]
In the classical applications of the $q$-analysis to mathematical physics the
background space is $\mathbb{R}$ or $\mathbb{C}$, these fields do not admit
$\left(  \mathbb{Z},+\right)  $ as a scale group. On these fields, by using
the Jackson derivative, it is possible to construct certain fractals, see
\cite{Ezran et al}. In the non-Archimedean setting, the background space is a
non-Archimedean local field, which has a fractal nature.

We introduce a non-Archimedean version of the Jackson derivative. Let $n$, $m$
be non-negative integers, and $f:$ $\mathbb{F}_{q}\left(  \left(  T\right)
\right)  \rightarrow\mathbb{C}$, we set
\[
\left(  \partial(n,m)f\right)  (x):=\frac{f(T^{-n}x)-f\left(  T^{m}x\right)
}{\left\vert T^{-n}x\right\vert -\left\vert T^{m}x\right\vert }=\frac
{f(T^{-n}x)-f\left(  T^{m}x\right)  }{\left(  q^{n}-q^{-m}\right)  \left\vert
x\right\vert }\text{ for }x\neq0\text{.}%
\]
This derivative measures the speed of deformation of function $f$ under the
scale group of $\mathbb{F}_{q}\left(  \left(  T\right)  \right)  $. Notice
that this derivative is\ not defined at the origin. In this article we use
mainly the case $\partial(1,1)=:\partial$. We also use the operators $\left(
q^{\pm N}f\right)  \left(  x\right)  :=f(T^{\mp1}x)$. The $q$-oscillator
algebras $\mathcal{A}_{q}$, $\mathcal{A}_{q}^{c}$ are generated by the symbols
$a$, $a^{\dag}$, $q^{\pm N}$. By interpreting $a$ as $\partial$, $a^{\dag}$ as
the multiplication $\left\vert x\right\vert $ and using $q^{\pm N}$, we show
the existence of a non-Archimedean realization of the Fock representation of
the algebras $\mathcal{A}_{q}$, $\mathcal{A}_{q}^{c}$. The underlying Hilbert
space of the representation is isometric to $L^{2}(\mathbb{K},dx)$, where $dx$
is the normalized Haar measure of $\left(  \mathbb{K},+\right)  $. This new
realization implies that many models constructed using $q$-oscillator algebras
are indeed non-Archimedean models, in particular, $p$-adic models.

In $q$-analysis, the parameter $q$ is a complex number, meanwhile in
$\mathbb{K}$-analysis, $q=p^{n}$, where $p$ is a prime number and $n$ is a
positive integer. By specializing the $q$ parameter to a power of $p$, we pass
from $q$-analysis to $\mathbb{K}$-analysis.

The $q$-deformed harmonic oscillators have been studied intensively, see,
e.g., \cite{Biedenharn}, \cite{Finkelstein-1999}, \cite{Lavagno-2010}%
-\cite{Lavagno et al 2}, \cite{Macfarlane}, \cite{Van der Jeugt}, among
others. These models can be formulated on $\mathbb{K}=\mathbb{F}_{q}\left(
\left(  T\right)  \right)  $. The energy levels of these harmonic oscillators
have the form%
\[
E_{n}=\frac{1}{2}\hslash\omega\frac{\sinh\left(  \frac{2n+1}{2}\ln q\right)
}{\sinh\left(  \frac{1}{2}\ln q\right)  }\sim\frac{1}{2}\hslash\omega
\frac{\exp\left(  \frac{2n+1}{2}\ln q\right)  }{\exp\left(  \frac{1}{2}\ln
q\right)  }\text{ as }n\rightarrow\infty\text{.}%
\]
These energy levels are no longer uniformly spaced since $q$ is a power of a
prime number. The interpretation of the non-uniform distribution of the energy
levels of the $q$-harmonic oscillator is a challenging problem. In the
non-Archimedean framework, they obey a scale law. Consider another background
space $\mathbb{K}_{m}=\mathbb{F}_{q^{m}}\left(  \left(  T\right)  \right)  $,
which is a $\mathbb{K}$-vector space of dimension $m\geq2$. In this new
background space, we have a copy of the $q$-deformed harmonic oscillator, with
energy levels:
\[
E_{n}^{\left(  m\right)  }=\frac{1}{2}\hslash\omega\frac{\sinh\left(
\frac{2n+1}{2}\ln q^{m}\right)  }{\sinh\left(  \frac{1}{2}\ln q^{m}\right)
}\sim\frac{1}{2}\hslash\omega\left(  \frac{\exp\left(  \frac{2n+1}{2}\ln
q\right)  }{\exp\left(  \frac{1}{2}\ln q\right)  }\right)  ^{m}.
\]
Then
\[
\frac{E_{n}^{\left(  m\right)  }}{\frac{1}{2}\hslash\omega}\sim\left(
\frac{E_{n}}{\frac{1}{2}\hslash\omega}\right)  ^{m}\text{ as }n\rightarrow
\infty\text{.}%
\]
This scale law is a reinterpretation of a well-known number-theoretic result,
which is available only in the non-Archimedean framework.

Let $V\left(  \left\vert x\right\vert \right)  :\mathbb{F}_{q}\left(  \left(
T\right)  \right)  \rightarrow\mathbb{R}$ be an arbitrary radial potential
with a unique singularity at the origin. In this article we propose the
following time-independent Schr\"{o}dinger equation:%
\[
\left\{
\begin{array}
[c]{l}%
\Psi_{n}:\mathbb{F}_{q}\left[  \left[  T\right]  \right]  \rightarrow
\mathbb{R}\\
\Psi_{n}\mid_{S}=0\\
\frac{-\hbar^{2}}{2m}\left\{  \left(  q^{-N}\partial\right)  ^{2}+V(\left\vert
x\right\vert )\right\}  \Psi_{n}(x)=E_{n}\Psi_{n}(x),
\end{array}
\right.
\]
here $\mathbb{F}_{q}\left[  \left[  T\right]  \right]  $ is the unit ball
centered at the origin. We show that the energy levels have the form
\begin{equation}
E_{n}=\frac{-\hbar^{2}}{2m}\left(  1-q^{-2}\right)  ^{2}q^{4n-4}%
+V(q^{-n})\text{, for \ }n=1,2,\ldots.\label{Energy_levels_A}%
\end{equation}
Here we determine only the point spectrum of $\frac{-\hbar^{2}}{2m}\left\{
\left(  q^{-N}\partial\right)  ^{2}+V(\left\vert x\right\vert )\right\}  $.
The determination of the whole spectrum is an open problem.

By a suitable selection of the potential $V\left(  \left\vert x\right\vert
\right)  $, the energy levels of several $q$-models can be obtained\ from
(\ref{Energy_levels_A}). For instance by taking,%
\[
V_{HA}\left(  \left\vert x\right\vert \right)  =\frac{\hbar^{2}\left(
1-q^{-2}\right)  ^{2}}{2mq^{2}\left\vert x\right\vert ^{4}}-\frac{1}{2}%
mc^{2}\left(  \frac{e^{2}}{\hbar c}\right)  ^{2}q^{4\mu}\frac{\left(
q-q^{-1}\right)  ^{2}}{\left(  \left\vert x\right\vert -\left\vert
x\right\vert ^{-1}\right)  ^{2}},\text{ }x\in\mathbb{F}_{q}\left[  \left[
T\right]  \right]  ,
\]
formula (\ref{Energy_levels_A}) gives the energy levels of the Finkelstein
$q$-hydrogen atom \cite{Finkelstein-1999}:%
\[
E_{n}(\mu)=-\frac{1}{2}mc^{2}\left(  \frac{e^{2}}{\hbar c}\right)  ^{2}%
\frac{q^{4\mu}}{\left[  2n+1\right]  ^{2}},
\]
where $\mu$\ is a real parameter, and $\left[  j\right]  =\frac{q^{j}%
-q^{-^{j}}}{q-q^{-1}}$. In the limit $q$ tends to one, (\ref{Energy_levels_A})
gives the Balmer energy formula \cite{Finkelstein-1999}.

The limits $\lim T\rightarrow1$, $\lim q\rightarrow1$ are completely
different. To the best of our knowledge, the understanding of the first one
requires motivic integration, while the second requires ordinary calculus, of
course, after extending the parameter $q$ as a real variable. Notice that
\ this difference is not very clear in the $p$-adic case, where $T$ is
replaced by $p$ and $q=p^{-1}$, for this reason, we prefer $\mathbb{F}%
_{q}\left(  \left(  T\right)  \right)  $ over $\mathbb{Q}_{p}$. It is
interesting to mention that in the limit $p$ tends to one, the $p$-adic
strings relate to ordinary strings see, e.g., \cite{Zuniga-Bocardo-Garcia-0}%
\ and the references therein.

The non-Archimedean difference equations introduced here are new mathematical
objects. There are several open problems and intriguing connections between
these $\pi$-difference equations with several mathematical theories.

\section{The $q$-oscillator algebras and Fock representations}

In this section we review some basic aspects of the $q$-oscillator
algebras$\ $(also called $q$-boson algebras) and their Fock representations.
For further details the reader may consult \cite[Chapter 5]{Klimyk}.

\subsection{The $q$-oscillator algebras}

Let $q$ be a fixed complex number such that $q\neq\pm1$. We recall that the
one-dimensional harmonic oscillator algebra is generated by two elements $a$,
$a^{\dagger}$ satisfying the commutation relation%
\begin{equation}
\left[  a,a^{\dagger}\right]  =aa^{\dagger}-a^{\dagger}a=1. \label{Formula_1A}%
\end{equation}
In the Fock representation the generators $a$, $a^{\dagger}$ correspond to the
annihilation and creation operators, respectively, and $N=a^{\dagger}a$
corresponds to the particle number operator. Furthermore,%
\begin{equation}
\left[  N,a^{\dagger}\right]  =a^{\dagger}\text{, }\left[  N,a\right]  =-a.
\label{Formula_1B}%
\end{equation}

\begin{definition}
The centrally extended $q$-oscillator algebra $\mathcal{A}_{q}^{c}$ is the
associative, unital, $\mathbb{C}$-algebra generated by four elements $a$,
$a^{\dagger}$, $q^{N}$, $q^{-N}$ subject to the relations{}

\begin{enumerate}
\item[1A.] $q^{-N}q^{N}=q^{N}q^{-N}=1$,

\item[2A.] $q^{N}a^{\dagger}=qa^{\dagger}q^{N}$,

\item[3A.] $q^{N}a=q^{-1}aq^{N}$,

\item[4A.] $\left[  a,a^{\dagger}\right]  _{q}:=aa^{\dagger}-qa^{\dagger
}a=q^{-N}$.
\end{enumerate}
\end{definition}

It is relevant to note that in the above definition $q^{N}$, $q^{-N}$, are
symbols for two elements and that $N$ is not an element of the algebra
$\mathcal{A}_{q}^{c}$.

For $\alpha\in\mathbb{C}$, $k\in\mathbb{Z}$, we use the notation%
\[
q^{kN+\alpha}:=q^{\alpha}\left(  q^{N}\right)  ^{k}\text{ and }\left[
N+\alpha\right]  :=\frac{q^{N+\alpha}-q^{-N-\alpha}}{q-q^{-1}}.
\]

\begin{definition}
\label{Definition_1}The symmetric $q$-oscillator algebra $\mathcal{A}_{q}$ is
the associative, unital, $\mathbb{C}$-algebra generated by four elements $a$,
$a^{\dagger}$, $q^{N}$, $q^{-N}$ subject to the relations{}

\begin{enumerate}
\item[1B.] $q^{-N}q^{N}=q^{N}q^{-N}=1$,

\item[2B.] $q^{N}a^{\dagger}=qa^{\dagger}q^{N}$,

\item[3B.] $q^{N}a=q^{-1}aq^{N}$,

\item[4B.] $\left[  a,a^{\dagger}\right]  _{q}:=aa^{\dagger}-qa^{\dagger
}a=q^{-N}$,

\item[5B.] $\left[  a,a^{\dagger}\right]  _{q^{-1}}:=aa^{\dagger}%
-q^{-1}a^{\dagger}a=q^{N}$.
\end{enumerate}
\end{definition}

Note that relations (4B)-(5B) imply that%
\[
a^{\dagger}a=\left[  N\right]  _{q}\text{, \ \ }aa^{\dagger}=\left[
N+1\right]  _{q},
\]
which in turn imply that%
\[
\left[  N\right]  _{q}a^{\dagger}=a^{\dagger}aa^{\dagger}=a^{\dagger}\left[
N+1\right]  _{q}\text{, and }a\left[  N\right]  _{q}=aa^{\dagger}a=\left[
N+1\right]  _{q}a.
\]
In the limit $q\rightarrow1$ the relations (1B)-(5B) of the algebra
$\mathcal{A}_{q}$ reduce to (\ref{Formula_1A}) and $N=a^{\dagger}a$.

\subsection{The Fock representation of $\mathcal{A}_{q}$}

In this section we assume that $q>0$, $q\neq1$. We do not use the bra and ket
notation, we directly identify the generators of $\mathcal{A}_{q}$ with
operators acting on a certain Hilbert space. This direct approach is more
convenient for our purposes, see \cite{Van der Jeugt}, \cite[Section
5.3]{Klimyk}

The Fock space $\mathcal{F}_{q}$ is an $\mathcal{A}_{q}$-module with basis
vectors
\[
v_{n},\text{ \ }n\in\mathbb{N}:=\left\{  0,1,\ldots,l,l+1,\ldots\right\}  ,
\]
and the action of the generators of $\mathcal{A}_{q}$ is given by
\[
q^{\pm N}v_{n}=q^{\pm n}v_{n}\text{, \ }a^{\dagger}v_{n}=\sqrt{[n+1]}%
v_{n+1}\text{, \ }av_{n}=\sqrt{[n]}v_{n-1},
\]
where%
\[
\lbrack r]:=(q^{r}-q^{-r})/(q-q^{-1})=\frac{\sinh(r\ln q)}{\sinh(\ln q)}.
\]
Notice that%
\[
a^{\dagger}av_{n}=[n]v_{n}\text{, \ }aa^{\dagger}v_{n}=[n+1]v_{n}.
\]
The Fock space becomes a Hilbert space with respect to the inner product
$\left\langle v_{m},v_{n}\right\rangle :=\delta_{m,n}$. Furthermore, $a$ and
$a^{\dagger}$ are adjoint to each other, whereas those of $q^{N}$ and $q^{-N}$
are self-adjoint operators.

\section{Non-Archimedean local fields}

We recall that the field of rational numbers $\mathbb{Q}$ admits two types of
norms: the Archimedean norm (the usual absolute value), and the
non-Archimedean norms (the $p$-adic norms) which are parameterized by the
prime numbers. The field of real numbers $\mathbb{R}$ arises as the completion
of $\mathbb{Q}$ with respect to the Archimedean norm. Fix a prime number $p$,
the $p$-adic norm is defined as
\[
\left\vert x\right\vert _{p}=\left\{
\begin{array}
[c]{lll}%
0 & \text{if} & x=0\\
p^{-\gamma} & \text{if} & x=p^{\gamma}\frac{a}{b}\text{,}%
\end{array}
\right.
\]
where $a$ and $b$ are integers coprime with $p$. The integer $ord(x):=\gamma$,
with $ord(0):=\infty$, is called the\textit{\ }$p$-\textit{adic order of} $x$.
The field of $p$-adic numbers $\mathbb{Q}_{p}$ is defined as the completion of
the field of rational numbers $\mathbb{Q}$ with respect to the $p$-adic norm
$|\cdot|_{p}.$

A non-Archimedean local field $\mathbb{K}$ is a locally compact topological
field with respect to a non-discrete topology, which comes from a norm
$\left\vert \cdot\right\vert _{\mathbb{K}}$ satisfying
\[
\left\vert x+y\right\vert _{\mathbb{K}}\leq\max\left\{  \left\vert
x\right\vert _{\mathbb{K}},\left\vert y\right\vert _{\mathbb{K}}\right\}  ,
\]
for $x,y\in\mathbb{K}$. Such a norm is called an \textit{ultranorm or
non-Archimedean}. Any non-Archimedean local field $\mathbb{K}$ of
characteristic zero is isomorphic (as a topological field) to a finite
extension of $\mathbb{Q}_{p}$. The field $\mathbb{Q}_{p}$ is the basic example
of non-Archimedean local field of characteristic zero. In the case of positive
characteristic, $\mathbb{K}$ is isomorphic to the field of formal Laurent
series $\mathbb{F}_{q}\left(  \left(  T\right)  \right)  $ over a finite field
$\mathbb{F}_{q}$, where $q$ is a power of a prime number $p$.

\begin{notation}
From now on, we fix the parameter $q$ to be a power of $p$. In addition. we
use $q$ to denote only the cardinality of $\mathbb{F}_{q}$.
\end{notation}

The \textit{ring of integers} of $\mathbb{K}$ is defined as
\[
R_{\mathbb{K}}=\left\{  x\in\mathbb{K};\left\vert x\right\vert _{\mathbb{K}%
}\leq1\right\}  .
\]
Geometrically $R_{\mathbb{K}}$ is the unit ball of the normed space $\left(
\mathbb{K},\left\vert \cdot\right\vert _{\mathbb{K}}\right)  $. This ring is a
domain of principal ideals having a unique maximal ideal, which is given by
\[
P_{\mathbb{K}}=\left\{  x\in\mathbb{K};\left\vert x\right\vert _{\mathbb{K}%
}<1\right\}  .
\]
We fix a generator $\pi$ of $P_{\mathbb{K}}$, i.e., $P_{\mathbb{K}}=\pi
R_{\mathbb{K}}$. Such a generator is also called a \textit{local uniformizing
parameter of} $\mathbb{K}$, and it plays the same role as $p$ in
$\mathbb{Q}_{p}.$

The \textit{group of units} of $R_{\mathbb{K}}$ is defined as
\[
R_{\mathbb{K}}^{\times}=\left\{  x\in R_{\mathbb{K}};\left\vert x\right\vert
_{\mathbb{K}}=1\right\}  .
\]
The natural map $R_{\mathbb{K}}\rightarrow R_{\mathbb{K}}/P_{\mathbb{K}}%
\cong\mathbb{F}_{q}$ is called the \textit{reduction mod} $P_{\mathbb{K}}$.
The quotient $\overline{\mathbb{K}}:=R_{\mathbb{K}}/P_{\mathbb{K}}%
\cong\mathbb{F}_{q}$, $q=p^{l}$, is called the \textit{residue field} of
$\mathbb{K}$. Every non-zero element $x$ of $\mathbb{K}$ can be written
uniquely as $x=\pi^{ord(x)}u$, $u\in R_{\mathbb{K}}^{\times}$. We call
$u:=ac\left(  x\right)  $ the \textit{angular component} of $x$ We set
$ord(0)=\infty$. The normalized valuation of $\mathbb{K}$ is the mapping
\[%
\begin{array}
[c]{ccc}%
\mathbb{K} & \rightarrow & \mathbb{Z}\cup\left\{  \infty\right\} \\
x & \rightarrow & ord(x).
\end{array}
\]
Then $\left\vert x\right\vert _{\mathbb{K}}=q^{-ord(x)}$ and $\left\vert
\pi\right\vert _{\mathbb{K}}=q^{-1}$.

We fix $\mathfrak{S}\subset R_{\mathbb{K}}$ a set of representatives of
$\mathbb{F}_{q}$ in $R_{\mathbb{K}}$, i.e., the reduction $\operatorname{mod}$
$P_{\mathbb{K}}$ is a bijection from $\mathfrak{S}$\ onto $\mathbb{F}_{q}$. We
assume that $0\in\mathfrak{S}$. Any non-zero element $x$ of $\mathbb{K}$ can
be written as
\begin{equation}
x=\pi^{ord(x)}\sum\limits_{i=0}^{\infty}x_{i}\pi^{i}, \label{Expansion}%
\end{equation}
where $x_{i}\in\mathfrak{S}$ and $x_{0}\neq0$. This series converges in the
norm $\left\vert \cdot\right\vert _{\mathbb{K}}$. Notice that $ac(x)=\sum
_{i=0}^{\infty}x_{i}\pi^{i}$.

A multiplicative character (or quasi-character) of the group $\left(
\mathbb{K}^{\times},\cdot\right)  $ is a continuous homomorphism
$\omega:\mathbb{K}^{\times}\rightarrow\mathbb{C}^{\times}$ satisfying
$\omega\left(  xy\right)  =\omega\left(  x\right)  \omega\left(  y\right)  $.
Every multiplicative character has the form%
\[
\omega\left(  x\right)  =\left\vert x\right\vert _{\mathbb{K}}^{s}\omega
_{0}\left(  ac(x)\right)  \text{, for some }s\in\mathbb{C}\text{,}%
\]
where $\omega_{0}$ is the restriction of $\omega$ to $R_{\mathbb{K}}^{\times}%
$; $\omega_{0}$ is a continuous multiplicative character of $\left(
R_{\mathbb{K}}^{\times},\cdot\right)  $ into the complex unit circle.

For an in-depth exposition of non-Archimedean local fields, the reader may
consult \cite{We,Taibleson}, see also \cite{Alberio et al,V-V-Z}.

\section{Non-Archimedean analogues of the Jackson derivative}

In this article we introduce several non-Archimedean analogues of the Jackson
derivative. For an in-depth presentation of the classical $q$-analysis the
reader may consult \cite{Ernst}, \cite{Kac}. Given $f:\mathbb{K}%
\rightarrow\mathbb{C}$ we define%
\[
\partial f\left(  x\right)  =\frac{f(\pi^{-1}x)-f\left(  \pi x\right)
}{\left(  q-q^{-1}\right)  \left\vert x\right\vert _{\mathbb{K}}}\text{, for
}x\neq0\text{.}%
\]
The existence of $\partial f\left(  0\right)  $ depends on $f$. In the
classical case, the value at the origin of the Jackson derivative is given by
the standard derivative, this approach cannot be used here.

If $g:\mathbb{K}\rightarrow\mathbb{C}$, then the following Leibniz-type rule
holds true:%
\begin{align}
\partial\left(  f\left(  x\right)  g(x)\right)   &  =g(\pi x)\partial f\left(
x\right)  +f(\pi^{-1}x)\partial g(x)\nonumber\\
&  =g(\pi^{-1}x)\partial f\left(  x\right)  +f(\pi x)\partial g(x).
\label{Leibniz_rule}%
\end{align}
Notice that for any function $f\left(  ac(x)\right)  $, it holds that
$\partial f\left(  ac(x)\right)  =0$, for $x\neq0$. In particular,
$\partial\omega_{0}\left(  ac(x)\right)  =0$, $x\neq0$, for any multiplicative
character $\omega_{0}$ of $\left(  R_{\mathbb{K}}^{\times},\cdot\right)  $. We
also have%
\begin{equation}
\partial\left\vert x\right\vert _{\mathbb{K}}^{m}=\left[  m\right]  \left\vert
x\right\vert _{\mathbb{K}}^{m-1}\text{ for }m\in\mathbb{N}\smallsetminus
\left\{  0\right\}  . \label{Formula_1}%
\end{equation}
We set $\left[  m\right]  !:=%
{\textstyle\prod\nolimits_{i=1}^{m}}
\left[  i\right]  $, with $\left[  0\right]  !=1$.

Another non-Archimedean Jackson-type derivative is defined as%
\[
\widetilde{\partial}f\left(  x\right)  =\frac{f\left(  \pi^{-1}x\right)
-f\left(  x\right)  }{\left(  q-1\right)  \left\vert x\right\vert
_{\mathbb{K}}},\text{ }x\neq0,
\]
where $f:\mathbb{K}\rightarrow\mathbb{C}$. Now, for $g:\mathbb{K}%
\rightarrow\mathbb{C}$, we have
\[
\widetilde{\partial}\left(  f\left(  x\right)  g\left(  x\right)  \right)
=g\left(  \pi^{-1}x\right)  \widetilde{\partial}f\left(  x\right)
+f(x)\widetilde{\partial}g\left(  x\right)  .
\]
Notice that
\begin{equation}
\widetilde{\partial}\left\vert x\right\vert _{\mathbb{K}}^{m}=\left[  \left[
m\right]  \right]  \left\vert x\right\vert _{\mathbb{K}}^{m-1},m\in
\mathbb{N\smallsetminus}\left\{  0\right\}  , \label{Formula_derivative}%
\end{equation}
where $\left[  \left[  m\right]  \right]  :=\frac{q^{m}-1}{q-1}$. We also set
$\left[  \left[  m\right]  \right]  !=%
{\textstyle\prod\nolimits_{i=1}^{m}}
\left[  \left[  i\right]  \right]  $, with $\left[  \left[  0\right]  \right]
!:=1$. In case of functions depending on several variables, say $f(x,t)$, we
use the notation $\partial_{x}f\left(  x,t\right)  $, $\widetilde{\partial
}_{x}f\left(  x,t\right)  $ to mean a derivative with respect to $x$.

\section{A non-Archimedean Fock representation of $\mathcal{A}_{q}$}

\subsection{Some operators}

We introduce the operators:
\[
a^{\dagger}f(x)=\left\vert x\right\vert _{\mathbb{K}}f(x)\text{,
\ }af(x)=\partial f\left(  x\right)  \text{, \ }q^{N}f\left(  x\right)
=f(\pi^{-1}x)\text{, }q^{-N}f\left(  x\right)  =f(\pi x),\text{\ }%
\]
which act on functions $f:\mathbb{K}\rightarrow\mathbb{C}$. We now fix a
multiplicative character $\omega_{vac}$ of $\left(  R_{\mathbb{K}}^{\times
},\cdot\right)  $. \ By simplicity we take $\omega_{vac}=1$. We call such a
function the\textit{ vacuum eigenstate}. We define%
\[
u_{n}\left(  x\right)  =\frac{\left\vert x\right\vert _{\mathbb{K}}^{n}}%
{\sqrt{\left[  n\right]  !}}\text{, }x\in\mathbb{K}\text{, for }n\in
\mathbb{N}\text{.}%
\]
Then%
\begin{equation}
a^{\dagger}u_{n}\left(  x\right)  =\sqrt{\left[  n+1\right]  }u_{n+1}\left(
x\right)  \text{ for }n\in\mathbb{N},\label{Formula_2}%
\end{equation}%
\begin{equation}
au_{n}\left(  x\right)  =\sqrt{\left[  n\right]  }u_{n-1}\left(  x\right)
\text{ for }n\in\mathbb{N\smallsetminus}\left\{  0\right\}  ,\label{Formula_3}%
\end{equation}%
\begin{equation}
au_{0}\left(  x\right)  =0,\label{Formula_4}%
\end{equation}%
\begin{equation}
q^{\pm N}u_{n}(x)=q^{\pm n}u_{n}(x)\text{ for }n\in\mathbb{N}\text{.}%
\label{Formula_5}%
\end{equation}

\subsection{A non-Archimedean Bargmann-Fock type realization}

Let $\mathcal{F}_{q}^{\blacklozenge}$ be the $\mathbb{C}$-vector space of
formal series of the form%
\[
f(x)=%
{\textstyle\sum\limits_{n=0}^{\infty}}
c_{n}\left\vert x\right\vert _{\mathbb{K}}^{n}\text{, }x\in\mathbb{K}\text{,
}c_{n}\in\mathbb{C}\text{ for every }n\text{.}%
\]
We introduce a sesquilinear form on $\mathcal{F}_{q}^{\blacklozenge}$ by
taking%
\[
\left(  f,g\right)  :=\overline{f\left(  \partial\right)  }g(x)\mid_{x=0},
\]
where $f(\partial):=\sum_{n=0}^{\infty}c_{n}\partial^{n}$. If $g(x)=\sum
_{n=0}^{\infty}d_{n}\left\vert x\right\vert _{\mathbb{K}}^{n}$, by using
(\ref{Formula_1}), we have%
\[
\left(  f,g\right)  =%
{\textstyle\sum\limits_{n=0}^{\infty}}
\overline{c_{n}}d_{n}\left[  n\right]  !.\text{ }%
\]
We set $\left\Vert f\right\Vert ^{2}:=%
{\textstyle\sum\limits_{n=0}^{\infty}}
\left\vert c_{n}\right\vert ^{2}\left[  n\right]  !$. Notice that
\[
\left(  u_{n}\left(  x\right)  ,u_{m}\left(  x\right)  \right)  =\delta_{n,m},
\]
i.e., $\left\{  u_{n}\left(  x\right)  \right\}  _{n\in\mathbb{N}}$ is an
orthonormal basis. We now set%
\[
\mathcal{F}_{q}:=\left\{  f=%
{\textstyle\sum\limits_{n=0}^{\infty}}
c_{n}u_{n}\left(  x\right)  \in\mathcal{F}_{q}^{\blacklozenge};\left\Vert
f\right\Vert ^{2}=%
{\textstyle\sum\limits_{n=0}^{\infty}}
\left\vert c_{n}\right\vert ^{2}<\infty\right\}  .
\]
Then, the space $\mathcal{F}_{q}$ endowed with the inner product $\left(
\cdot,\cdot\right)  $ becomes a Hilbert space having $\left\{  u_{n}\left(
x\right)  \right\}  _{n\in\mathbb{N}}$ as an orthonormal basis. Also,
$\mathcal{F}_{q}$ is isomorphic to the classical $l^{2}\left(  \mathbb{C}%
\right)  $ Hilbert space consisting of the square-summable complex sequences.

By formulae (\ref{Formula_2})-(\ref{Formula_5}), the operators $a^{\dagger}$,
$a$,\ $q^{N}$, $q^{-N}$ are well-defined \ in the $\mathbb{C}$-vector space%
\[
\mathcal{F}_{q}^{\text{fin}}:=\left\{  f(x)=%
{\textstyle\sum\limits_{n=0}^{M}}
c_{n}u_{n}\left(  x\right)  \in\mathcal{F}_{q}\text{; for some }%
M=M(f)\right\}  ,
\]
which is a dense subspace of $\mathcal{F}_{q}$. These operators extend to
linear unbounded operators on $\mathcal{F}_{q}$. One easily verifies that
$a^{\dagger}$ is the adjoint of $a$ and that $q^{N}$, $q^{-N}$ are
self-adjoint operators on $\mathcal{F}_{q}^{\text{fin}}$.

Finally, by using formulae (\ref{Formula_2})-(\ref{Formula_5}), one verifies
that the operators $a^{\dagger}$, $a$,\ $q^{N}$, $q^{-N}$ satisfy the
relations (1)-(5) given in Definition \ref{Definition_1}. A similar
realization for the algebra $\mathcal{A}_{q}^{c}$ exists.

\subsection{$L^{2}$ is isometric to $\mathcal{F}_{q}$}

We fix a Haar measure $dx$ on the additive group $\left(  \mathbb{K},+\right)
$ satisfying that $\int_{R_{\mathbb{K}}}dx=1$. The space $L^{2}\left(
\mathbb{K}\right)  $ consists of all the functions $f:\mathbb{K}%
\rightarrow\mathbb{C}$ satisfying
\[
\left\Vert f\right\Vert _{2}^{2}=%
{\displaystyle\int\limits_{\mathbb{K}}}
\left\vert f(x)\right\vert ^{2}dx<\infty\text{.}%
\]
In the case $\mathbb{K}=\mathbb{Q}_{p}$ it is well-known that $L^{2}\left(
\mathbb{Q}_{p}\right)  $ admits a countable orthonormal basis, see, e.g.,
\cite{KKZuniga}, \cite{Koch}, \cite{V-V-Z}. Consequently $L^{2}\left(
\mathbb{Q}_{p}\right)  $ is isometric to $l^{2}\left(  \mathbb{C}\right)  $.
This fact is indeed valid for any non-Archimedean local field. We set
\[
\omega_{rbk}\left(  x\right)  :=q^{\frac{-r}{2}}\chi_{_{\mathbb{K}}}\left(
\pi^{-1}k\left(  \pi^{r}x-b\right)  \right)  \Omega\left(  \left\vert \pi
^{r}x-b\right\vert _{\mathbb{K}}\right)  ,
\]
where $r\in\mathbb{Z}$, $k\in\mathfrak{S}\smallsetminus\left\{  0\right\}  $,
$b\in\mathbb{K}/R_{\mathbb{K}}$, $b=\sum_{i=\beta}^{-1}n_{i}\pi^{i}$, with
$n_{i}\in\mathfrak{S}$, $\beta\in\mathbb{Z}$, $\beta<0$ ($\mathfrak{S}$ is a
set of representatives of $\mathbb{F}_{q}$ in $R_{\mathbb{K}}$),
$\chi_{_{\mathbb{K}}}$ is the standard additive character of the additive
group $\left(  \mathbb{K},+\right)  $, i.e., $\chi_{_{\mathbb{K}}}%
:\mathbb{K}\rightarrow\mathbb{C}$ is a continuous mapping satisfying:
$\left\vert \chi_{_{\mathbb{K}}}\left(  x\right)  \right\vert =1$,
$\chi_{_{\mathbb{K}}}\left(  x+y\right)  =\chi_{_{\mathbb{K}}}\left(
x\right)  \chi_{_{\mathbb{K}}}\left(  y\right)  $, $\chi_{_{\mathbb{K}}}%
\mid_{R_{\mathbb{K}}}=1$ but $\chi_{_{\mathbb{K}}}\mid_{\mathbb{K}%
\smallsetminus R_{\mathbb{K}}}\neq1$. Finally, $\Omega\left(  \left\vert
\pi^{r}x-b\right\vert _{\mathbb{K}}\right)  $ denotes the characteristic
function of the ball $\pi^{-r}b+\pi^{-r}R_{\mathbb{K}}$. The family $\left\{
\omega_{rbk}\left(  x\right)  \right\}  _{rbk}$ forms a complete orthonormal
basis of $L^{2}\left(  \mathbb{K}\right)  $. The proof of this result follows
using the argument of the case $\mathbb{K}=\mathbb{Q}_{p}$, see \cite[Theorem
2]{Kozyrev}. Therefore $L^{2}\left(  \mathbb{K}\right)  $ is isometric to
$l^{2}\left(  \mathbb{C}\right)  $, the Fock space.

\begin{remark}
(i) Given $f$, $g$, $\partial f$, $\partial g\in L^{2}\left(  \mathbb{K}%
\right)  $, by using changes of variables, one verifies that%
\[
\left\langle g,\partial f\right\rangle :=%
{\displaystyle\int\limits_{\mathbb{K}}}
\overline{g\left(  x\right)  }\text{ }\partial f\left(  x\right)  \text{ }dx=-%
{\displaystyle\int\limits_{\mathbb{K}}}
\overline{\partial g\left(  x\right)  }\text{ }f\left(  x\right)  \text{ }dx.
\]

\noindent(ii) The operators $\left\vert x\right\vert _{\mathbb{K}}$,
$\partial$ are well-defined on the space of test functions which is dense in
$L^{2}\left(  \mathbb{K}\right)  $. However these operators cannot be directly
interpreted as creation and annihilation operators in $L^{2}\left(
\mathbb{K}\right)  $. Let $\mathfrak{i}:L^{2}\left(  \mathbb{K}\right)
\rightarrow$ $l^{2}\left(  \mathbb{C}\right)  $ be the above mentioned
isometry. Then the creation operator, respectively annihilation operator, in
$L^{2}\left(  \mathbb{K}\right)  $ are $\mathfrak{i}^{-1}\circ\left\vert
x\right\vert _{\mathbb{K}}\circ\mathfrak{i}$, respectively $\mathfrak{i}%
^{-1}\circ\partial\circ\mathfrak{i}$.
\end{remark}

\section{The non-Archimedean Harmonic Oscillator}

Motivated by Biedenharn's work \cite{Biedenharn}, see also \cite{Klimyk}\ and
the references therein, we introduce the $\pi$-momentum operator
$\boldsymbol{\Pi}$ and the $\pi$-position operator $\boldsymbol{Q}$, in terms
of $a^{\dagger}=\left\vert x\right\vert _{K}$, $a\boldsymbol{=}\partial$, as%
\[
\boldsymbol{\Pi}:=i\sqrt{\frac{m\hslash\omega}{2}}\left(  a^{\dagger
}-a\right)  \text{, \ }\boldsymbol{Q}:=\sqrt{\frac{\hslash}{2m\omega}}\left(
a^{\dagger}+a\right)  .
\]
The Hamiltonian of the $\pi$-harmonic oscillator (or non-Archimedean
oscillator) is defined as%
\[
\boldsymbol{H}=\frac{\boldsymbol{\Pi}^{2}}{2m}+\frac{m\omega^{2}}%
{2}\boldsymbol{Q}^{2}=\frac{1}{2}\hslash\omega\left(  aa^{\dagger}+a^{\dagger
}a\right)  .
\]
Then $\boldsymbol{H}u_{n}\left(  x\right)  =E_{n}u_{n}\left(  x\right)  $,
i.e., $\boldsymbol{H}$ is diagonal on the eigenstates $u_{n}\left(  x\right)
$ with eigenvalues%
\[
E_{n}:=\frac{1}{2}\hslash\omega\left(  \left[  n+1\right]  +\left[  n\right]
\right)  =\frac{1}{2}\hslash\omega\frac{\sinh\left(  \frac{2n+1}{2}\ln
q\right)  }{\sinh\left(  \frac{1}{2}\ln q\right)  }.
\]
Since $q$ is a power of a prime number, the energy levels are no longer
uniformly spaced. In the limit $q\rightarrow1$ these numbers give the
eigenvalues of the usual quantum harmonic oscillator, see, e.g.,
\cite{Biedenharn}. The interpretation of the non-uniform distribution of the
energy levels of the $q$-harmonic oscillator is a challenging problem.

In the non-Archimedean framework, the nonuniform spacing of the energy levels
of the $\pi$-harmonic oscillator obey to a scale law. We set $S_{r}=\left\{
x\in\mathbb{K};\left\vert x\right\vert _{\mathbb{K}}=q^{r}\right\}  $ for the
sphere with center at the origin and radius $r\in\mathbb{Z}$. Notice that
$S_{0}=%
{\textstyle\bigsqcup\nolimits_{j}}
\left(  j+\pi R_{\mathbb{K}}\right)  $, where $j\in\mathfrak{S}\smallsetminus
\left\{  0\right\}  $, see (\ref{Expansion}), and each ball $j+\pi
R_{\mathbb{K}}$ can be identified with an infinite regular rooted tree. The
regularity means that each vertex has exactly $q$ children. The set
$\mathbb{K\smallsetminus}\left\{  0\right\}  $ is the disjoint union of a
countable number of copies of scaled trees. More precisely,%
\[
\mathbb{K\smallsetminus}\left\{  0\right\}  =%
{\displaystyle\bigsqcup\limits_{r=-\infty}^{\infty}}
S_{r}=%
{\displaystyle\bigsqcup\limits_{r=-\infty}^{\infty}}
\pi^{-r}S_{0}.
\]
The group $(\mathbb{Z},+)$ is a scale group acting on $\mathbb{K\smallsetminus
}\left\{  0\right\}  $ as
\[%
\begin{array}
[c]{lll}%
\mathbb{Z\times}\left(  \mathbb{K\smallsetminus}\left\{  0\right\}  \right)
& \rightarrow & \mathbb{K\smallsetminus}\left\{  0\right\}  \\
\left(  r,x\right)   & \rightarrow & \pi^{-r}x.
\end{array}
\]
Then $\mathbb{K\smallsetminus}\left\{  0\right\}  $ is a self-similar set
obtained from $S_{0}$ by the action of the scale group $(\mathbb{Z},+)$.

Let us fix $\mathbb{K}$ a non-Archimedean local field, which plays the role of
a one-dimensional background space. Now, let $\mathbb{K}_{m}$ be an extension
of $\mathbb{K}$\ of degree $m$, this means that $\mathbb{K}_{m}$\ is a local
field, containing $\mathbb{K}$, which is a $\mathbb{K}$-vector space of
dimension $m\geq2$. The ring of integers $R_{\mathbb{K}}$ of $\mathbb{K}$ is a
subring of the ring of integers $R_{\mathbb{K}_{m}}$ of $\mathbb{K}_{m}$. The
local uniformizing parameter $\pi$ of $\mathbb{K}$ generates an ideal $\pi
R_{\mathbb{K}_{m}}$ in $R_{\mathbb{K}_{m}}$. Since any ideal of $R_{\mathbb{K}%
_{m}}$ has the form $\pi_{m}^{l}R_{\mathbb{K}_{m}}$, where $\pi_{m}$ denotes a
local uniformizing parameter of $\mathbb{K}_{m}$, we have $\pi R_{\mathbb{K}%
_{m}}=\pi_{m}^{e}R_{\mathbb{K}_{m}}$, for some positive integer $e$ (called
\textit{the ramification index} of the extension $\mathbb{K}/\mathbb{K}_{m}$).
By a well-known result, we have $m=ef$, where the positive integer $f$ (called
\textit{the inertia index} of the extension $\mathbb{K}/\mathbb{K}_{m}$) is
the dimension of $\overline{\mathbb{K}_{m}}$\ considered as a $\overline
{\mathbb{K}}$-vector space, i.e., $\overline{\mathbb{K}_{m}}=\mathbb{F}%
_{q^{f}}$, see, e.g., \cite{We}.

Since any function $f:\mathbb{K}_{m}\rightarrow\mathbb{C}$ has a restriction
to $\mathbb{K}$, the operators $a_{m}^{\dagger}=\left\vert x\right\vert
_{\mathbb{K}_{m}}$, $a_{m}\boldsymbol{=}\partial$ have natural restrictions
which act on functions defined on $\mathbb{K}$. We denote these restrictions
as $a^{\dagger}$, $a$. Then, we may assume the existence of `two identical
copies' of a non-Archimedean harmonic oscillator, one in $\mathbb{K}$ and the
other in $\mathbb{K}_{m}$. The energy levels of these oscillators are%
\begin{equation}
E_{n}\left(  \mathbb{K}\right)  =\frac{1}{2}\hslash\omega\frac{\sinh\left(
\frac{2n+1}{2}\ln q\right)  }{\sinh\left(  \frac{1}{2}\ln q\right)  }\text{,
\ \ }E_{n}\left(  \mathbb{K}_{m}\right)  =\frac{1}{2}\hslash\omega\frac
{\sinh\left(  \frac{2n+1}{2}\ln q^{f}\right)  }{\sinh\left(  \frac{1}{2}\ln
q^{f}\right)  },\label{Energy_Formulas}%
\end{equation}
respectively.

By using (\ref{Energy_Formulas}),
\[
E_{n}\left(  \mathbb{K}\right)  \sim\frac{1}{2}\hslash\omega\frac{\exp\left(
\frac{2n+1}{2}\ln q\right)  }{\exp\left(  \frac{1}{2}\ln q\right)  }\text{,
\ \ }E_{n}\left(  \mathbb{K}_{m}\right)  \sim\frac{1}{2}\hslash\omega\left[
\frac{\exp\left(  \frac{2n+1}{2}\ln q\right)  }{\exp\left(  \frac{1}{2}\ln
q\right)  }\right]  ^{f}%
\]
for $n\rightarrow\infty$, then%
\[
\left(  \frac{E_{n}\left(  \mathbb{K}_{m}\right)  }{\frac{1}{2}\hslash\omega
}\right)  \sim\left(  \frac{E_{n}\left(  \mathbb{K}\right)  }{\frac{1}%
{2}\hslash\omega}\right)  ^{f}\text{ for }n\rightarrow\infty.
\]

\section{Non-Archimedean quantum mechanics}

In this section we introduce a new class of $q$-deformed Schr\"{o}dinger
equations and study the cases of the free particle and a particle in a
non-Archimedean box.

\subsection{A non-Archimedean Heisenberg uncertainty relations}

In the algebra $\mathcal{A}_{q}$, it verifies that $\partial\left\vert
x\right\vert _{\mathbb{K}}-q^{-1}\left\vert x\right\vert _{\mathbb{K}}%
\partial\boldsymbol{=}q^{N}$, and by using $q^{-1}q^{-N}\left\vert
x\right\vert _{\mathbb{K}}=\left\vert x\right\vert _{\mathbb{K}}q^{-N}$, we
have%
\begin{align}
1  &  =q^{-N}\partial\left\vert x\right\vert _{\mathbb{K}}-q^{-1}%
q^{-N}\left\vert x\right\vert _{\mathbb{K}}\partial=q^{-N}\partial\left\vert
x\right\vert _{\mathbb{K}}-q^{-2}\left\vert x\right\vert _{\mathbb{K}}%
q^{-N}\partial\nonumber\\
&  =:\left[  q^{-N}\partial,\left\vert x\right\vert _{\mathbb{K}}\right]
_{q^{-2}}. \label{CCR}%
\end{align}
We propose using operator $-i\hslash q^{-N}\partial$ as a non-Archimedean
analogue of the momentum operator, and propose operator $\left\vert
x\right\vert _{\mathbb{K}}$ as an analogue of the position operator. By using
(\ref{CCR}), the Heisenberg uncertainty formula becomes%
\begin{equation}
\left[  -i\hslash q^{-N}\partial,\left\vert x\right\vert _{\mathbb{K}}\right]
_{q^{-2}}=-i\hslash. \label{Heisenberg}%
\end{equation}
The relation (\ref{Heisenberg}) is a $q$-deformation of the classical
Heisenberg uncertainty relation. In the limit $q\rightarrow1$ the relation
(\ref{Heisenberg}) becomes the standard Heisenberg uncertainty relation.

\subsection{Some mathematical results}

We review some results on $q$-analysis following \cite{Ernst}, \cite[Chapter
2]{Klimyk}, \cite{Kac}. In this framework $q$ is a complex parameter, since
here $q$ represents the cardinality of a finite field, we use a
complex\ parameter $\rho$ in our review of the $\rho$-analysis, later on we
specialize $\rho$\ to $q$.

For any nonzero complex number $\rho$, the $\rho$-number $\left[  a\right]
_{\rho}$, $a\in\mathbb{C}$, is defined as%
\[
\left[  a\right]  _{\rho}=\frac{\rho^{a}-\rho^{-a}}{\rho-\rho^{-1}}%
=\frac{\sinh\left(  a\ln\rho\right)  }{\sinh\left(  \ln\rho\right)  }.
\]
We also define%
\[
\left[  \left[  a\right]  \right]  _{\rho}=\frac{\rho^{a}-1}{\rho-1}%
=\rho^{\frac{a\left(  a-1\right)  }{2}}\left[  a\right]  _{\sqrt{\rho}}.
\]
In the case $\rho=q$, we use the simplified notation $\left[  a\right]
_{q}=\left[  a\right]  $, $\left[  \left[  a\right]  \right]  _{q}=\left[
\left[  a\right]  \right]  $. We use this convention for any function
depending on $\rho$.

For $m\in\mathbb{N}$, we set $\rho$-factorial $\left[  m\right]  _{\rho}!:=%
{\textstyle\prod\nolimits_{j=1}^{m}}
\left[  m\right]  _{\rho}$ with $\left[  0\right]  _{\rho}!:=1$, and $\left[
\left[  m\right]  \right]  _{\rho}!:=%
{\textstyle\prod\nolimits_{j=1}^{m}}
\left[  \left[  m\right]  \right]  _{\rho}$ with $\left[  \left[  0\right]
\right]  _{\rho}!:=1$. By convention, $\left[  m\right]  _{q}!=\left[
m\right]  !$, $\left[  \left[  m\right]  \right]  _{q}!=\left[  \left[
m\right]  \right]  !$.

We also set for $m\in\mathbb{N}$,%
\[
\left(  a;\rho\right)  _{m}:=\left\{
\begin{array}
[c]{lll}%
\left(  1-a\right)  \left(  1-a\rho\right)  \cdots\left(  1-a\rho^{m-1}\right)
& \text{for} & m\geq1\\
1 & \text{for} & m=0.
\end{array}
\right.
\]
Then
\begin{equation}
\left[  m\right]  _{\rho}!=\frac{\rho^{-\frac{m\left(  m-1\right)  }{2}}%
}{\left(  1-\rho^{2}\right)  ^{m}}\left(  \rho^{2};\rho^{2}\right)  _{m}.
\label{factorial_fomula}%
\end{equation}
For $\left\vert \rho\right\vert <1$, we set%
\[
\left(  a;\rho\right)  _{\infty}:=%
{\displaystyle\prod\limits_{j=1}^{\infty}}
\left(  1-a\rho^{j-1}\right)  .
\]
This product converges for all $a\in\mathbb{C}$ and defines an analytic function.

\subsubsection{The $\rho$-exponential functions}

There are two $\rho$-analogues of the exponential function:%
\[
e_{\rho}\left(  z\right)  :=%
{\displaystyle\sum\limits_{n=0}^{\infty}}
\frac{z^{n}}{\left(  \rho;\rho\right)  _{n}}=\frac{1}{\left(  z;\rho\right)
_{\infty}}\text{ for }z,\rho\in\mathbb{C}\text{, with }\left\vert
\rho\right\vert <1\text{,}%
\]%
\begin{equation}
\mathcal{E}_{\rho}\left(  z\right)  :=%
{\displaystyle\sum\limits_{n=0}^{\infty}}
\frac{\rho^{\frac{n\left(  n-1\right)  }{2}}z^{n}}{\left(  \rho;\rho\right)
_{n}}=\left(  -z;\rho\right)  _{\infty}\text{ for }z,\rho\in\mathbb{C}\text{,
with }\left\vert \rho\right\vert <1\text{.}\label{Formula_Buztos}%
\end{equation}
Furthermore, $e_{\rho}\left(  z\right)  \mathcal{E}_{\rho}\left(  z\right)
=1$,
\begin{align*}
e_{\rho}\left(  z\right)   &  :=1+%
{\displaystyle\sum\limits_{n=1}^{\infty}}
\frac{\left(  \frac{z}{1-\rho}\right)  ^{n}}{\left(  1-\rho\right)  \left(
1-\rho^{2}\right)  \cdots\left(  1-\rho^{n}\right)  }=%
{\displaystyle\sum\limits_{n=1}^{\infty}}
\frac{\left(  \frac{z}{1-\rho}\right)  ^{n}}{\left[  \left[  n\right]
\right]  _{\rho}!}\\
&  =\frac{1}{%
{\displaystyle\prod\limits_{j=1}^{\infty}}
\left(  1-\frac{z}{1-\rho}\rho^{j-1}\right)  },
\end{align*}
and also%
\[
\mathcal{E}_{\rho}\left(  z\right)  =%
{\displaystyle\sum\limits_{n=0}^{\infty}}
\frac{\rho^{\frac{n\left(  n-1\right)  }{2}}\left(  \frac{z}{1-\rho}\right)
^{n}}{\left[  \left[  n\right]  \right]  _{\rho}!}=%
{\displaystyle\prod\limits_{j=1}^{\infty}}
\left(  1-\frac{z}{1-\rho}\rho^{j-1}\right)  ,
\]%
\begin{equation}
\mathcal{E}_{\rho^{2}}\left(  z\right)  =%
{\displaystyle\sum\limits_{n=0}^{\infty}}
\frac{\rho^{\frac{n\left(  n-1\right)  }{2}}\left(  \frac{z}{1-\rho^{2}%
}\right)  ^{n}}{\left[  n\right]  _{\rho}!}=%
{\displaystyle\prod\limits_{j=1}^{\infty}}
\left(  1+\frac{z}{1-\rho^{2}}\rho^{j-1}\right)  .\label{Expo_2}%
\end{equation}

\subsubsection{The $\pi$-exponential functions}

For $\lambda\in\mathbb{C}$ and $x\in\mathbb{K}$, by using $\rho=q^{-1}$ in
(\ref{Expo_2}) and $\left[  n\right]  _{\rho}=\left[  n\right]  _{^{\rho^{-1}%
}}$, we set%
\begin{equation}
\mathcal{E}\left(  x,\lambda\right)  :=%
{\displaystyle\sum\limits_{l=0}^{\infty}}
\frac{q^{\frac{-\left(  l-1\right)  l}{2}}\lambda^{l}\left\vert x\right\vert
_{\mathbb{K}}^{l}}{\left[  l\right]  !}=\mathcal{E}_{q^{-2}}\left(
\frac{\lambda\left\vert x\right\vert _{\mathbb{K}}}{1-q^{-2}}\right)  =%
{\displaystyle\prod\limits_{j=1}^{\infty}}
\left(  1+\frac{\lambda\left\vert x\right\vert _{\mathbb{K}}}{1-q^{-2}%
}q^{-2j+2}\right)  ,\label{Formula_E_entire}%
\end{equation}
and
\begin{equation}
e\left(  x,\lambda\right)  :=%
{\displaystyle\sum\limits_{l=0}^{\infty}}
\frac{\lambda^{l}\left\vert x\right\vert _{\mathbb{K}}^{l}}{\left[  \left[
l\right]  \right]  _{q^{-1}}!}=e_{\rho}\left(  \left(  1-q^{-1}\right)
\lambda\left\vert x\right\vert _{\mathbb{K}}\right)  =\frac{1}{%
{\displaystyle\prod\limits_{j=1}^{\infty}}
\left(  1-\left(  1-q^{-1}\right)  \lambda\left\vert x\right\vert
_{\mathbb{K}}q^{-j+1}\right)  }.\label{Formula_e_meromorphic}%
\end{equation}
Notice that the series $e\left(  x,\lambda\right)  $\ converges for
$\left\vert \lambda\right\vert \left\vert x\right\vert _{\mathbb{K}}<\frac
{1}{1-q^{-1}}$, and that $\mathcal{E}\left(  x,\lambda\right)  $\ and
$e\left(  x,\lambda\right)  $\ are radial functions of $x$. We call $e\left(
x;\lambda\right)  $, $\mathcal{E}\left(  x,\lambda\right)  $ the $\pi
$\textit{-exponential functions}.

By using (\ref{Formula_derivative}), we have%
\begin{equation}
\widetilde{\partial}_{x}^{m}e\left(  x;\lambda\right)  =\lambda^{m}e\left(
x,\lambda\right)  \text{, for }m\in\mathbb{N\smallsetminus}\left\{  0\right\}
. \label{plane-wave_e}%
\end{equation}
On the other hand, $\partial_{x}\mathcal{E}\left(  x,\lambda\right)
=\lambda\mathcal{E}\left(  \pi^{-1}x,\lambda\right)  =\lambda q^{N}%
\mathcal{E}\left(  x,\lambda\right)  $, i.e.,
\[
\left(  q^{-N}\partial_{x}\right)  \mathcal{E}\left(  x,\lambda\right)
=\lambda\mathcal{E}\left(  x,\lambda\right)  .
\]
By induction on $m$,
\begin{equation}
\left(  q^{-N}\partial_{x}\right)  ^{m}\mathcal{E}\left(  x;\lambda\right)
=\lambda^{m}\mathcal{E}\left(  x,\lambda\right)  \text{, for }m\in
\mathbb{N\smallsetminus}\left\{  0\right\}  . \label{Eigenfunctions}%
\end{equation}

\subsubsection{$\rho$-Trigonometric functions}

The $\rho$-trigonometric functions attached to $\mathcal{E}_{\rho}\left(
z\right)  $ are defined as%
\[
\sin_{\rho}\left(  z\right)  =\frac{1}{2i}\left(  \mathcal{E}_{\rho}\left(
iz\right)  -\mathcal{E}_{\rho}\left(  -iz\right)  \right)  =%
{\displaystyle\sum\limits_{n=0}^{\infty}}
\frac{\left(  -1\right)  ^{n}\rho^{\left(  2n+1\right)  n}\left(  \frac
{z}{1-\rho}\right)  ^{2n+1}}{\left[  \left[  2n+1\right]  \right]  _{\rho}!},
\]%
\[
\cos_{\rho}\left(  z\right)  =\frac{1}{2}\left(  \mathcal{E}_{\rho}\left(
iz\right)  +\mathcal{E}_{\rho}\left(  -iz\right)  \right)  =%
{\displaystyle\sum\limits_{n=0}^{\infty}}
\frac{\left(  -1\right)  ^{n}\rho^{\left(  2n-1\right)  n}\left(  \frac
{z}{1-\rho}\right)  ^{2n}}{\left[  \left[  2n\right]  \right]  _{\rho}!},
\]
where $i=\sqrt{-1}$.

We define the $\pi$-trigonometric functions attached to $\mathcal{E}\left(
x,\lambda\right)  $ as%
\begin{align*}
\cos\left(  x,\mu\right)   &  :=%
{\displaystyle\sum\limits_{n=0}^{\infty}}
\frac{\left(  -1\right)  ^{n}q^{n\left(  2n-1\right)  }\mu^{2n}\left\vert
x\right\vert _{\mathbb{K}}^{2n}}{\left[  2n\right]  !}\text{, }\\
\sin\left(  x,\mu\right)   &  :=%
{\displaystyle\sum\limits_{n=0}^{\infty}}
\frac{\left(  -1\right)  ^{n}q^{n\left(  2n+1\right)  }\mu^{2n+1}\left\vert
x\right\vert _{\mathbb{K}}^{2n+1}}{\left[  2n+1\right]  !},
\end{align*}
for $x\in\mathbb{K}$, $\mu\in\mathbb{R}$. We call $\cos\left(  x,\mu\right)
$, resp. $\sin\left(  x;\mu\right)  $, the $\pi-$\textit{cosine function},
resp., the $\pi-$\textit{sine function}. Some useful properties of these
trigonometric functions are the following:%
\begin{align*}
\cos\left(  0,\mu\right)   &  =1\text{, }\cos\left(  x,-\mu\right)
=\cos\left(  x,\mu\right)  \text{, }q^{-N}\partial\cos\left(  x,\mu\right)
=-\mu\sin\left(  x,\mu\right)  ,\\
\sin\left(  0,\mu\right)   &  =0\text{, }\sin\left(  x,-\mu\right)
=-\sin\left(  x,-\mu\right)  \text{, }q^{-N}\partial\sin\left(  x,\mu\right)
=\mu\cos\left(  x,\mu\right)  .
\end{align*}

\subsection{A non-Archimedean analogue of the Schr\"{o}dinger equation}

Given a function $\Psi(t,x):\mathbb{K}\times\mathbb{K\rightarrow C}$, we set%
\[
\partial_{t}u(t,x):=\frac{u(\pi^{-1}t,x)-u(\pi t,x)}{\left(  q-q^{-1}\right)
\left\vert t\right\vert _{\mathbb{K}}},\text{ \ }q_{x}^{-N}\partial
_{x}u(t,x):=\frac{u(t,x)-u(t,\pi^{2}x)}{\left(  1-q^{-2}\right)  \left\vert
x\right\vert _{\mathbb{K}}}.
\]
We propose the following non-Archimedean analogue of the Schr\"{o}dinger
equation:%
\begin{equation}
i\hbar\partial_{t}\Psi(t,x)=\left\{  \frac{-\hbar^{2}}{2m}\left(  q_{x}%
^{-N}\partial_{x}\right)  ^{2}+V(t,x)\right\}  \Psi(t,x),
\label{Schrodinger-Equation}%
\end{equation}
where $\Psi(t,x)$ is a wave function, $m$ is the mass of the particle, and
$V(t,x)$ is the potential. The time and spatial variables are elements of a
non-Archimedean local field of arbitrary characteristic, but the wave
functions are complex-valued.

\subsection{Free particle}

In the case of the free particle, the time-independent Schr\"{o}dinger
equation takes the form%
\begin{equation}
\frac{-\hbar^{2}}{2m}\left(  q_{x}^{-N}\partial_{x}\right)  ^{2}\Psi
(x)=E\Psi(x). \label{Eq_1}%
\end{equation}
We study the solutions of (\ref{Eq_1}) in spaces of type $\mathcal{F}%
_{q}^{\blacklozenge}$. We first notice that if $\Psi(x)$ is a solution of
(\ref{Eq_1}), then $h(ac(x))\Psi(x)$ is also a solution of (\ref{Eq_1}), where
$h(ac(x))$ is an arbitrary function. This means that a regularization at the
origin for $\Psi(x)$\ is required. By using that $\left(  q_{x}^{-N}%
\partial_{x}\right)  ^{2}\mathcal{E}\left(  x;\lambda\right)  =\lambda
^{2}\mathcal{E}\left(  x,\lambda\right)  $, we obtain that
\begin{align}
\Psi(x)  &  =c_{0}\mathcal{E}\left(  x,i\sqrt{\frac{2mE}{\hslash^{2}}}\right)
+c_{1}\mathcal{E}\left(  x,-i\sqrt{\frac{2mE}{\hslash^{2}}}\right) \nonumber\\
&  =a_{0}\cos\left(  x,\sqrt{\frac{2mE}{\hslash^{2}}}\right)  +a_{1}%
\sin\left(  x,\sqrt{\frac{2mE}{\hslash^{2}}}\right)  \label{Solution}%
\end{align}
where $a_{0}$, $a_{1}$ are functions depending only on the angular component
of $x$. Since the \ angular component is not defined at the origin, function
$\Psi(x)$ is not defined at the origin. Since $\cos\left(  0,\sqrt{\frac
{2mE}{\hslash^{2}}}\right)  =1$, $\sin\left(  0,\sqrt{\frac{2mE}{\hslash^{2}}%
}\right)  =0$, the simplest way of regularizing $\Psi(x)$ at the origin is by
choosing $a_{0}=0$ and $a_{1}$ as a nonzero constant, i.e., by choosing%
\begin{equation}
\Psi(x)=a_{1}\sin\left(  x,\sqrt{\frac{2mE}{\hslash^{2}}}\right)  .
\label{Sol_E_Sch}%
\end{equation}
The functions $\mathcal{E}\left(  x,\pm i\sqrt{\frac{2mE}{\hslash^{2}}%
}\right)  $ are the analogs of the classical plane waves. These functions are
radial functions defined in $\mathbb{K}$ by a convergent, complex-valued
series, see (\ref{Formula_E_entire}).

Another possible version of a non-Archimedean, $q$-deformed, time-independent
Schr\"{o}dinger equation\ is
\begin{equation}
\frac{-\hbar^{2}}{2m}\widetilde{\partial}_{x}^{2}\Psi(x)=E\Psi(x).
\label{Schrodinger-Equation-2}%
\end{equation}
By (\ref{plane-wave_e}), the planes waves have the form $e\left(  x,\pm
i\sqrt{\frac{2mE}{\hslash^{2}}}\right)  $ are solutions of
(\ref{Schrodinger-Equation-2}). These series converge when $\left\vert
x\right\vert _{\mathbb{K}}\sqrt{\frac{2mE}{\hslash^{2}}}<\frac{1}{1-q^{-1}}$,
see (\ref{Formula_e_meromorphic}). For this reason we focus on equations of
type (\ref{Eq_1}).

\subsection{Particle in a box}

We now consider a potential of the form%
\[
V(x)=\left\{
\begin{array}
[c]{lll}%
0 & \text{if} & \left\vert x\right\vert _{\mathbb{K}}\leq q^{L-1}\\
\infty & \text{if} & \left\vert x\right\vert _{\mathbb{K}}>q^{L},
\end{array}
\right.
\]
where $L$ is a fixed integer. We look for a solution of the Schr\"{o}dinger
equation
\[
\left\{  \frac{-\hbar^{2}}{2m}\left(  q_{x}^{-N}\partial_{x}\right)
^{2}+V(x)\right\}  \Psi(x)=E\Psi(x)
\]
subjected to the conditions: $\Psi(0)=0$, $\Psi(q^{L})=0$. The first is a
regularization condition and the second condition guarantees that the particle
is confined in the box $\left\{  x\in K;\left\vert x\right\vert _{\mathbb{K}%
}\leq q^{L-1}\right\}  $. The first condition implies that the solution has
the form (\ref{Sol_E_Sch}), with $E=\frac{\hbar^{2}}{2m}\mu^{2}$, $\mu
\in\mathbb{R}$. Notice that $\Psi(x)=\Psi(\left\vert x\right\vert
_{\mathbb{K}})$. To satisfy the second condition, we need
\[
\Psi(q^{L})=B_{1}\sin\left(  q^{L},\sqrt{\frac{2mE}{\hslash^{2}}}\right)
=0\text{.}%
\]
There is a sequence of positive real numbers $\omega_{1}<\omega_{2}%
<\cdots<\omega_{k}\cdots$, such that\ the zeros of $\sin\left(  x,\mu\right)
=0$\ are $\left\vert x\right\vert _{\mathbb{K}}\mu=\omega_{k}$, $k\geq1$, then
$\mu=q^{-L}\omega_{k}$, $k\geq1$. This fact follows from Theorem 5.1 in
\cite{Bustoz et al}. Indeed, the exponential function in \cite[formula
2.2]{Bustoz et al}\ is exactly $\mathcal{E}(\left(  1-\rho^{2}\right)  z)$,
with $q=\rho^{2}$. Therefore%
\[
E_{k}=\frac{\hbar^{2}\omega_{k}^{2}}{2mq^{2L}}\text{, for }k\geq1,
\]
are the energy levels for a particle confined in a non-Archimedean box.

\section{The non-Archimedean Schr\"{o}dinger equation with a radial potential}

The ball $B_{L}$ with center at the origin and radius $L\in\mathbb{Z}$ is
defined as
\[
B_{L}=\left\{  x\in\mathbb{K};\left\vert x\right\vert _{\mathbb{K}}\leq
q^{L}\right\}  .
\]
The sphere $S_{L}$\ with center at the origin and radius $L\in\mathbb{Z}$ is
defined as
\[
S_{L}=\left\{  x\in\mathbb{K};\left\vert x\right\vert _{\mathbb{K}}%
=q^{L}\right\}  .
\]
We denote by $\left[  -\infty,+\infty\right]  $ the extended numeric line. We
fix a function $V:\left[  0,+\infty\right)  \rightarrow\left[  -\infty
,+\infty\right]  $. We denote by $Sing(V)$ the set of singularities of $V$. A
point $x\in\left[  0,+\infty\right)  $\ is a singular point of $V$, if $V$ is
not continuous at $x$, or if $V(x)=\pm\infty$. \ The function $V\left(
\left\vert x\right\vert _{K}\right)  $ is a radial potential. We now consider
the following eigenvalue problem:%
\begin{equation}
\left\{
\begin{array}
[c]{l}%
\Psi:B_{L}\rightarrow\mathbb{C}\\
\left\{  \frac{-\hbar^{2}}{2m}\left(  q_{x}^{-N}\partial_{x}\right)
^{2}+V(\left\vert x\right\vert _{\mathbb{K}})\right\}  \Psi(x)=E\Psi(x).
\end{array}
\right.  \label{Problem_2}%
\end{equation}
By using (\ref{Eigenfunctions}) with $m=2$, and taking $\Psi(x)=\mathcal{E}%
\left(  \left\vert x\right\vert _{\mathbb{K}},\lambda\right)  $, one gets
\[
\left(  q^{-N}\partial_{x}\right)  ^{2}\mathcal{E}\left(  \left\vert
x\right\vert _{\mathbb{K}};\lambda\right)  =\lambda^{2}\mathcal{E}\left(
\left\vert x\right\vert _{\mathbb{K}},\lambda\right)  .
\]
Now, by replacing $\Psi(x)$ in (\ref{Problem_2}), we obtain the condition%
\begin{equation}
\left\{  \frac{-\hbar^{2}}{2m}\lambda^{2}+V(\left\vert x\right\vert
_{\mathbb{K}})-E)\right\}  \mathcal{E}\left(  \left\vert x\right\vert
_{\mathbb{K}},\lambda\right)  =0.\label{Eq-10}%
\end{equation}
Consider the points $\left\vert x\right\vert _{\mathbb{K}}=q^{l}\leq q^{L}$
such that $x\notin Sing(V)$. Then (\ref{Eq-10}) becomes%
\begin{equation}
\left\{  \frac{-\hbar^{2}}{2m}\lambda^{2}+V(q^{l})-E)\right\}  \mathcal{E}%
\left(  q^{l},\lambda\right)  =0.\label{Eq-11}%
\end{equation}
By (\ref{Formula_E_entire}) the zeros of $\mathcal{E}\left(  q^{l}%
,\lambda\right)  =0$ satisfy $1+\frac{\lambda q^{l}}{1-q^{-2}}q^{-2j+2}=0$,
for $j=1,2,\ldots,$ therefore, the energy levels $E=E_{l,\lambda}$ have the
form%
\[
E_{l,\lambda}=\frac{-\hbar^{2}}{2m}\lambda^{2}+V(q^{l})\text{, }%
\]
where $l$ is an integer satisfying $l\leq L$ such that $S_{l}\nsubseteqq
Sing(V)$, and
\[
\lambda\in\mathbb{R}\smallsetminus\left\{  -\left(  1-q^{-2}\right)
q^{2j-l-2};j\geq1\right\}  .
\]

\subsection{Potentials supported in the unit ball}

To obtain a more precise description of the energy levels, it is necessary to
impose boundary conditions and some additional restrictions to the function
$V$.

We take
\[
V(\left\vert x\right\vert _{\mathbb{K}}):B_{0}\rightarrow\left[
-\infty,+\infty\right]  ,
\]
such that $Sing(V)$ is just the origin. We consider the following eigenvalue
problem:%
\begin{equation}
\left\{
\begin{array}
[c]{l}%
\Psi:B_{0}\rightarrow\mathbb{C}\\
\Psi\mid_{S_{0}}=0\\
\left\{  \frac{-\hbar^{2}}{2m}\left(  q_{x}^{-N}\partial_{x}\right)
^{2}+V(\left\vert x\right\vert _{\mathbb{K}})\right\}  \Psi(x)=E\Psi(x).
\end{array}
\right.  \label{Problem_3}%
\end{equation}
We take $\Psi(x)=\mathcal{E}\left(  \left\vert x\right\vert _{\mathbb{K}%
},\lambda\right)  $. To satisfy the condition $\Psi\mid_{S_{0}}=0$, we
require
\[
1+\frac{\lambda}{1-q^{-2}}q^{-2j+2}=0\text{, for \ }j=1,2,\ldots\text{,}%
\]
i.e.,
\begin{equation}
\lambda=-\left(  1-q^{-2}\right)  q^{2j-2}\text{, for \ }j=1,2,\ldots
\text{.}\label{Eq-12}%
\end{equation}
Take $\left\vert x\right\vert _{\mathbb{K}}=q^{-r}<1$, notice that $x\notin
Sing(V)$, then (\ref{Eq-11}), with $-r=l$, is satisfied if \ $\lambda$
satisfies (\ref{Eq-12}). We pick $j=r$, then the energy levels are given by
\begin{equation}
E_{r}=\frac{-\hbar^{2}}{2m}\left(  1-q^{-2}\right)  ^{2}q^{4r-4}%
+V(q^{-r})\text{, for \ }r=1,2,\ldots\text{,}\label{Energy_Levels_A}%
\end{equation}
and the functions
\[
\Psi_{r}(x)=\mathcal{E}\left(  \left\vert x\right\vert _{\mathbb{K}},-\left(
1-q^{-2}\right)  q^{2r-2}\right)  \text{ for \ }r=1,2,\ldots\text{,}%
\]
are eigenfunctions. The determination of all the possible eigenfunctions
requires solving an equation of the form%
\[
\frac{-\hbar^{2}}{2m}\left(  \frac{1-q^{-2}}{q^{2}}\right)  ^{2}y^{4}%
+V(y^{-1})=E,
\]
for $y\in\left(  0,1\right)  $, where $E$ is known, and then take
$r=\frac{-\ln y}{\ln q}\in\mathbb{N}.$

\subsection{$\pi-$Hydrogen atom}

By a suitable selection of the potential $V(\left\vert x\right\vert
_{\mathbb{K}})$, the energy levels of several $q$-models can be obtained\ from
(\ref{Energy_Levels_A}). For instance by taking,%
\begin{align*}
V_{HO}\left(  \left\vert x\right\vert _{\mathbb{K}}\right)   &  =\frac
{\hbar^{2}\left(  1-q^{-2}\right)  ^{2}}{2mq^{2}\left\vert x\right\vert
_{\mathbb{K}}^{4}}+\frac{1}{2}\hslash\omega\frac{\sinh\left(  \ln q^{\frac
{1}{2}}\left\vert x\right\vert \right)  }{\sinh\left(  \frac{1}{2}\ln
q\right)  }\\
&  =\frac{\hbar^{2}\left(  1-q^{-2}\right)  ^{2}}{2mq^{2}\left\vert
x\right\vert _{\mathbb{K}}^{4}}+\frac{1}{2}\hslash\omega\frac{q^{\frac{1}{2}%
}\left\vert x\right\vert -q^{\frac{-1}{2}}\left\vert x\right\vert ^{-1}%
}{q^{\frac{1}{2}}-q^{\frac{-1}{2}}},
\end{align*}
formula (\ref{Energy_Levels_A}) gives the energy levels of the $q$-harmonic
oscillator, see \cite{Biedenharn}.

Many versions of the $q$-hydrogen atom have been studied. In
\cite{Finkelstein-1999}, Finkelstein studied a model of a $q$-hydrogen atom
with energy levels of the form%
\begin{equation}
E_{n}(\mu)=-\frac{1}{2}mc^{2}\left(  \frac{e^{2}}{\hbar c}\right)  ^{2}%
\frac{q^{4\mu}}{\left[  2n+1\right]  ^{2}}, \label{Energy_levels_HA}%
\end{equation}
where $\mu$\ is a real parameter. This result was established using classical
$q$-analysis on $\mathbb{C}$. The potential%
\begin{align*}
V_{HA}\left(  \left\vert x\right\vert \right)   &  =\frac{\hbar^{2}\left(
1-q^{-2}\right)  ^{2}}{2mq^{2}\left\vert x\right\vert _{\mathbb{K}}^{4}}%
-\frac{1}{2}mc^{2}\left(  \frac{e^{2}}{\hbar c}\right)  ^{2}q^{4\mu}%
\frac{\sinh^{2}\left(  \ln q\right)  }{\sinh^{2}\left(  \ln\left\vert
x\right\vert \right)  }\\
&  =\frac{\hbar^{2}\left(  1-q^{-2}\right)  ^{2}}{2mq^{2}\left\vert
x\right\vert _{\mathbb{K}}^{4}}-\frac{1}{2}mc^{2}\left(  \frac{e^{2}}{\hbar
c}\right)  ^{2}q^{4\mu}\frac{\left(  q-q^{-1}\right)  ^{2}}{\left(  \left\vert
x\right\vert -\left\vert x\right\vert ^{-1}\right)  ^{2}}%
\end{align*}
produces the energy levels (\ref{Energy_levels_HA}). In the limit $q$ tends to
one, (\ref{Energy_levels_HA}) becomes%
\[
E=-\frac{1}{2}mc^{2}\left(  \frac{e^{2}}{\hbar c}\right)  ^{2}\frac{1}{\left(
2n+1\right)  ^{2}},
\]
which is the Balmer energy formula, where the principal quantum number is
$2n+1$, see \cite{Finkelstein-1999}.

\section{Some open problems}

The construction of non-Archimedean quantum mechanics as a $q$-deformation of
the classical quantum mechanics gives rise to several new mathematical
problems and intriguing connections.

\subsection{Semigroups with non-Archimedean time}

A central problem is to determine if there is a semigroup attached to
Schr\"{o}dinger \ equation (\ref{Schrodinger-Equation}), i.e., if there is a
family of operators \ $\left\{  S_{t}\right\}  _{t\in\mathbb{K}}$ such that
\[
\Psi(t,x)=S_{t}\Psi_{0}\left(  x\right)  \text{, with }\Psi(0,x)=\Psi
_{0}\left(  x\right)  :\mathbb{K}\rightarrow\mathbb{K},
\]
is the solution of the initial valued problem attached to
(\ref{Schrodinger-Equation}).

\subsection{A non-Archimedean version of the Frobenius method}

Set $\boldsymbol{D}=\left(  q^{-N}\partial\right)  $ and $A_{i}(x)=\sum
_{l=0}^{\infty}\frac{c_{i,l}\left\vert x\right\vert _{\mathbb{K}}^{l}}{\left[
l\right]  !}$ for $i=0,1,\ldots,M$. To determine if a $\pi$-difference
equation of the form%
\begin{equation}
\sum_{i=1}^{M}A_{i}(x)\boldsymbol{D}^{i}\Phi\left(  x\right)  =0
\label{p-adic-Equations}%
\end{equation}
admits a solution of the from $\Phi\left(  x\right)  =\sum_{l=0}^{\infty}%
\frac{d_{l}\left\vert x\right\vert _{\mathbb{K}}^{l+\gamma}}{\left[  l\right]
!}:\mathbb{K}\rightarrow\mathbb{C}$. To the best of our knowledge, there is no
a theory for equations of type (\ref{p-adic-Equations}). It is important to
mention here, that nowadays there are at least three different types of
theories of $p-$adic differential equations, see \cite{Alberio et al},
\cite{Buibuim}, \cite{Kedlaya}, \cite{KKZuniga}, \cite{Koch},
\cite{Ralph-Simanaca}, and \cite{Zuniga-LNM-2016}.

\subsection{Non-Archimedean representations of $q$-oscillatory algebras}

Suppose that $g:\mathbb{K}\rightarrow\mathbb{K}$. We define the operators%
\[
\Delta g(x)=\frac{g\left(  \pi^{-1}x\right)  -g\left(  \pi x\right)  }{\left(
\pi^{-1}-\pi\right)  x}\text{, for }x\neq0\text{,}%
\]
and%
\[
\widetilde{\Delta}g(x)=\frac{g\left(  \pi^{-1}x\right)  -g\left(  x\right)
}{\left(  \pi^{-1}-1\right)  x}\text{, for }x\neq0\text{.}%
\]
Is it possible to construct a Fock-type representation \ of $\mathcal{A}_{q}$,
where $\mathfrak{a}g=\Delta f$ and $\mathfrak{a}^{\dagger}g=xg$? \ A solution
of this problem will allow constructing non-Archimedean quantum mechanics with
$\mathbb{K}$-valued wave functions via quantum groups.

Another relevant problem is to study $\pi$-difference equations of type%
\begin{equation}%
{\displaystyle\sum\limits_{j=1}^{L}}
a_{j}\left(  x\right)  \Delta^{j}g(x)=0, \label{p-adic-Equations-2}%
\end{equation}
where $a_{j}\left(  x\right)  =\sum_{k=0}^{\infty}d_{j,k}x^{k}$ with $d_{j,k}%
$s$\in\mathbb{K}$, and $g:\mathbb{K}\rightarrow\mathbb{K}$. Notice that
equations of type (\ref{p-adic-Equations}) are radically different to those of
type (\ref{p-adic-Equations-2}).

\subsection{Sato-Bernstein-type theorems}

A very relevant problem consists in studying the existence of Sato-Bernstein
theorems on algebras of type $\mathbb{C}\left[  \left\vert x\right\vert
_{\mathbb{K}},\partial,q^{-N},q^{N}\right]  $, see, e.g., \cite{Bjork}.

\end{document}